\documentclass[]{aastex631}

\usepackage{amsmath}
\usepackage{amssymb}
\usepackage{nicefrac}
\usepackage{booktabs}
\usepackage{soul}
\usepackage{xcolor}
\usepackage{natbib}

\newcommand{\hmpc}{\ensuremath{h\text{Mpc}^{-1}}}

\hypersetup{citecolor=blue}
\begin{document}

\title{$\nu$GAN: A Generative Adversarial Emulator for Cosmic Web with Neutrinos}

\correspondingauthor{Neerav Kaushal}
\email{kaushal@mtu.edu}

\author[0000-0003-4786-2348]{Neerav Kaushal}
\affiliation{Institute of Computing and Cybersystems, Michigan Technological University, Houghton, MI 49931, USA}

\author[0000-0003-3052-3059]{Elena Giusarma}
\affiliation{Department of Physics, Michigan Technological University, Houghton, MI 49931, USA}
              
\author{Mauricio Reyes}
\affiliation{Department of Physics, Michigan Technological University, Houghton, MI 49931, USA}

\begin{abstract}

Understanding the impact of neutrino masses on the evolution of Universe is a crucial aspect of modern cosmology. Due to their large free streaming lengths, neutrinos significantly influence the formation of cosmic structures at non-linear scales. To maximize the information yield from current and future galaxy surveys, it is essential to generate precise theoretical predictions of structure formation. One approach to achieve this is by running large sets of cosmological numerical simulations, which is a computationally intensive process. 
In this study, we propose a deep learning-based generative adversarial network (GAN) model to emulate the Universe for a variety of neutrino masses. Our model called $\nu$GAN (for neutrino GAN) is able to generate 2D cosmic webs of the Universe for a number of neutrino masses ranging from 0.0 eV to 0.4 eV. The generated maps exhibit statistical independence, lack correlations with training data, and very closely resemble the distribution of matter in true maps. We assess the accuracy of our results both visually and through key statistics used in cosmology and computer vision analyses. Our results indicate that samples generated by $\nu$GAN are accurate within a 5\% error on power spectrum between $k=0.01$ to $k=0.5$~\hmpc. 
Although this accuracy covers the mildly non-linear scales, consistent with other works and observations, achieving higher accuracy at fully non-linear scales requires more sophisticated models, such as diffusion models. Nevertheless, our work opens up new avenues for building emulators to generate fast and massive neutrino simulations, potentially revolutionizing cosmological predictions and analyses. This work serves as a proof-of-concept, paving the way for future extensions with higher-resolution 3D data and advanced generative models. 

\end{abstract}


\section{Introduction} \label{sec:intro}

Neutrinos are among the most abundant particles in the Universe, with number densities only slightly lower than those of photons. In the early Universe, neutrinos were relativistic, contributing to the radiation energy density during that epoch. Unlike photons, neutrinos have now become non-relativistic and possess rest mass, thereby contributing to the total matter density of the Universe. This indicates that relic neutrinos can have significant effects on cosmological observables, particularly influencing the background evolution, the spectra of matter perturbations, and the anisotropies in the Cosmic Microwave Background (CMB).

Next-generation large-scale structure (LSS) surveys, including Dark Energy Spectroscopic Instrument (DESI)~\citep{DESI}, eROSITA~\citep{erosita}, Euclid~\citep{euclid}, Nancy Grace Roman Space Telescope (WFIRST)~\citep{wfirst}, and Rubin Observatory's Legacy Survey of Space and Time (LSST)~\citep{lsst}, will map vast cosmological volumes with high precision, producing datasets that span a wide range of redshifts and cosmic epochs. These surveys are expected to significantly improve constraints on cosmological parameters, including the sum of neutrino masses, thereby enhancing our understanding of the fundamental role neutrinos play in the evolution of the Universe. Recent analyses from DESI and the Atacama Cosmology Telescope (ACT)~\citep{DESI:2025, ACT:2025} have already refined cosmological constraints on $\sum m_\nu$. Current upper bounds from joint analyses of CMB and LSS data lie in the range $\sum m_\nu < 0.06$–$0.08$ eV at the 95\% confidence level, assuming a minimal $\Lambda$CDM+$\sum m_\nu$ framework.

Extracting the full scientific potential of these upcoming surveys requires theoretical predictions of comparable precision. In particular, accurate comparison between observations and theory demands rigorous modeling of the spatial distribution of matter and luminous tracers, especially in scenarios involving massive neutrinos. Analytic tools such as perturbation theory~\citep{Bernardeau_review}, can provide reliable predictions on  quasi-linear scales, \cite[e.g.][]{Quijote, Samushia_2021, Gualdi_2021, Kuruvilla_2021, Bayer_2021,  Banerjee_2019, Changhoon_2019, Uhlemann_2020, Friedrich_2020, Massara_2020, Dai_2020, Allys_2020, Banerjee_2020, Banerjee_2021, Gualdi_2020, Giri_2020, Bella_2020, Changhoon_2020, Valgiannis_2021,  Kuruvilla_2021b}, but most of the cosmological information resides in the fully non-linear regime, which requires more sophisticated approaches. In the absence of a comprehensive analytical model, numerical simulations with massive neutrinos are crucial for studying these non-linear scales and validating observations. In particular, N-body simulations are among the most crucial tools for evolving cosmological matter fluctuations under gravity alone, allowing for direct comparisons with theoretical predictions. These simulations are instrumental in generating mock galaxy catalogs, computing covariance matrices, and optimizing observational strategies. Recent developments in N-body simulations incorporating massive neutrinos have significantly advanced our understanding of their impact on clustering at fully non-linear scales in real space \citep{bird, lesgourgues, Villaescusa_2013, Paco2014, peloso}, the clustering and abundance of halos and cosmic voids \citep{Ema2014, DEMNUni_castorina, massara_2015}, and matter clustering in real space \citep{paco_neutrino_rsd, bel_2019}. However, a significant drawback of cosmological N-body simulations is their high computational cost. A single simulation demands extensive computational resources and can take days or even weeks to complete. This computational bottleneck restricts the amount of information that can be extracted from observational data and tested against theoretical models. Consequently, there is a pressing need for faster methods to generate cosmological simulations that maintain both accuracy and reliability, thereby accelerating the process and enhancing our ability to extract and analyze cosmological information. 

Over the past decade, deep learning has emerged as a powerful tool for emulating high-resolution numerical simulations and accelerating cosmological predictions. Convolutional neural networks (CNNs) have been widely used for parameter inference, weak lensing map generation, and super-resolution of N-body simulations~\citep{mudur2, diff1, Rodr_guez_2018, Siyu_2018, Renan_2020, necola, recon23, recon24, Li_etal, Ni_etal, rcroft2, rcroft1, andreas1, adam1}. CNNs have also been successfully applied to model the impact of massive neutrinos on structure formation~\citep{Giusarma1, Giusarma:2019feb}, demonstrating their effectiveness in extracting cosmological information from complex datasets. These models can efficiently learn to emulate the evolution of cosmic structures from high-fidelity simulations, significantly reducing computational costs.

More recently, generative models such as Generative Adversarial Networks (GANs)~\citep{Rodr_guez_2018, gan_winther, gan_perraudin1, Andrianomena:2022ucu}, normalizing flows\citep{sultan2, Lovell_2021}, and diffusion models~\citep{mudur2, Mudur, Mudur_2025, Sether:2024ltp} have demonstrated remarkable success in generating synthetic cosmological data that closely match high-fidelity simulations. While CNNs and generative models have been extensively applied to a range of cosmological tasks, GAN-based emulators specifically tailored for massive neutrino cosmologies remain relatively unexplored. Our work aims to fill this gap by developing a GAN framework conditioned on neutrino mass, enabling rapid generation of cosmic web realizations across different neutrino mass scenarios. This approach builds on the success of generative models for cosmic structure emulation while addressing the unique challenges posed by massive neutrino effects on non-linear structure formation.

In particular, we employ deep GANs to generate 2D cosmic web realizations of the Universe, conditioned on a range of neutrino masses. We assume a scenario of three degenerate neutrinos, where the electron, muon, and tau neutrino species share the same mass. The choice of GANs is motivated by their ability to effectively learn the complex probability distributions underlying the data, enabling the generation of new, random, statistically independent, and identically distributed samples after training on N-body simulations. These generated samples are uncorrelated with the training examples. Our model, named $\nu$GAN, is specifically conditioned on neutrino masses, allowing it to produce dark matter cosmic webs for any given neutrino mass after training. This approach substantially reduces the computational burden associated with generating variable-mass neutrino simulations using traditional methods, as numerous new samples can be generated within seconds.

To validate our results, we employ various summary statistics, including the power spectrum, transfer function, pixel intensity histograms, peak statistics, and structural similarity tests.

This study serves as a proof of concept, demonstrating that generative models, such as GANs, are a feasible and effective approach to emulate cosmological simulations with massive neutrinos. Building on this foundation, future work will extend the framework to more advanced generative architectures, including diffusion models and normalizing flows, which promise improved accuracy, better coverage of the data distribution, and greater fidelity on non-linear scales.

This paper is organized as follows. Section \ref{sec:methods_cgan} briefly introduces and discusses conditional GAN and the data we used for training our GAN model. Section \ref{sec:implementation_cgan} details the training process, model architecture, and hyperparameters while section \ref{sec:results_cgan} presents the quantitative results. Finally, we draw our conclusions in section \ref{sec:conclusions_and_discussions_cgan}.

\section{Methods} \label{sec:methods_cgan}

\subsection{Conditional GAN}\label{subsec:cgan2}

A conditional Generative Adversarial Network or CGAN~\citep{gan} is an extension of the traditional GAN framework, designed to generate data samples with specific characteristics or attributes. In a standard GAN, two neural networks, a generator and a discriminator, are employed. The generator produces synthetic data that mimics real data, while the discriminator distinguishes between real and generated data. These networks compete during training: the generator improves its ability to create realistic data, and the discriminator enhances its skill in differentiating real from fake data. Both networks undergo simultaneous training, and if the optimization process is executed effectively, the generator will acquire the ability to generate data that matches the data distribution.
In a CGAN, an additional conditional element, often in the form of extra input data or labels, is introduced. This conditional information is used to guide the generation process, enabling the model to generate data samples with specific attributes or characteristics. 

In this study, we condition both the generator ($G$) and the discriminator ($D$) on a parameter denoted as $y$, representing the neutrino mass. By doing so, we ensure that the generated data samples are specifically tailored to different neutrino masses.

The algorithm works as follows:

\begin{enumerate}
        \item The input consists of a random noise vector, denoted as $z$, which can originate from various distributions such as the Gaussian distribution (typically sampled from a unit-normal distribution, $N(0,1)$), a uniform distribution, or other structured inputs.
	
        \item The generator $G$, which is parameterized by a neural network, takes the latent vector $z$ and an additional random variable $y$ (serving as a conditioning factor), and produces the output $G(z,y)$.
        
	  \item The discriminator $D$, also parameterized by a neural network, takes real data samples $x$ and the synthetic samples generated by $G$, named $G(z,y)$. It then provides scores for both, denoted as $D(x)$ and $D(G(z,y))$ respectively. These scores represents the discriminator's confidence in whether the samples are real, i.e. originating from the actual data distribution $p_{data}(x)$. When scaled to the range $[0,1]$, this score can also be loosely interpreted as an implicit likelihood of the data given $D$, i.e., $p(x|D)$.
 	
	\item The predictions made by $D$ are compared to the actual, true labels, and a loss is computed, represented as $L(D,G)$.
	
	\item This loss is backpropagated through $D$ and then through $G$, to update the weights and biases of both networks.
	
	\item These steps, from 1 to 5, are iteratively repeated over multiple epochs while processing the entire dataset.
\end{enumerate}

\subsection{Data} \label{subsec:data}
In the following, we present a detailed description of the various steps involved in the generation and preprocessing of data:

\begin{enumerate}
    \item We run N-body simulations using the COLA (COmoving Lagrangian Acceleration) approximation~\citep{cola, cola2} with the MG-PICOLA code~\footnote{https://github.com/HAWinther/MG-PICOLA-PUBLIC}~\citep{mgpicola}. COLA is an approximate simulation method designed to efficiently evolve large-scale structure by combining Lagrangian perturbation theory with N-body dynamics, achieving faster computation while preserving large-scale accuracy. We used MG-PICOLA instead of full N-body simulations because it provides a computationally efficient compromise, retaining accuracy on large and mildly non-linear scales at a fraction of the computational cost. The simulations track the evolution of $1024^3$ cold dark matter (CDM) particles in the presence of neutrinos with masses $0.0$, $0.1$, and $0.4$ eV from a redshift of $z=9$ to $z=0$ in $50$ timesteps in a simulation volume of size $500$ Mpc/h. The cosmological parameters used for these simulations are $\Omega_M = 0.3175$, $\Omega_B = 0.0490$, $n_s = 0.9624$, $\sigma_8 = 0.8340$, $H = 67.11$, and $\sum m_\nu = (0.0, 0.1, 0.4)$~eV. We generate two realizations for each neutrino mass, resulting in a total of six realizations. Each simulation provides the 3D spatial coordinates of $1024^3$ CDM particles.

    \item Following \cite{Rodr_guez_2018}, for each realization, we take this 3D cube of particle positions and divide the positions along the x-axis into 1000 equal segments. We then extract 2D slices of particle positions in the y-z plane and select 500 non-consecutive 2D slices. We repeat this process along the y and z axes, obtaining 1500 2D slices per realization, and a total of 9000 slices for the entire dataset (i.e., six realizations).

    \item  These 2D slices are then pixelized into \(256 \times 256\) images, where each pixel value represents the number of particles within that pixel. This process effectively transforms the data into 2D grayscale images of size \(256 \times 256\). The images are then smoothed using a Gaussian filter with a standard deviation of 1, resulting in floating-point pixel values spanning a wide range of magnitudes, rather than discrete integers. As part of data augmentation, each image is randomly flipped horizontally, vertically, or both, each with a 25\% probability; the remaining 25\% of the time, the image is left unaltered.
    
    \item The data is scaled to $[-1,1]$, following~\citep{Rodr_guez_2018}, using the same scaling parameter $a=4$ as in that work. This scaling improves model performance and ensures compatibility with the generator’s final \texttt{tanh} activation. The original data ($\rho$) and the scaled data ($\rho(x)$) are related by the following transformation:
    
    \begin{equation}
        \rho(x) = \frac{2x}{x+a} - 1~.
        \label{eqn:scaling}
    \end{equation}
    
\end{enumerate}
    This transformation is analogous to the logarithmic function. As the cosmic web of the Universe spans a dynamic range of magnitudes, from nearly empty cosmic voids to super-massive galaxy clusters, this transformation enhances the contrast of filaments, galaxy sheets, and dark matter halos. The parameter $a$ in equation \ref{eqn:scaling} controls the median value of the images. The choice $a=4$ balances compression of extreme densities while retaining sufficient contrast in intermediate regions, ensuring the transformed data are centered around zero and compatible with the generator’s \texttt{tanh} activation, promoting stable learning.

\section{Implementation} \label{sec:implementation_cgan}

Learning the optimal parameters for the discriminator and generator networks, and effectively optimizing a GAN, are challenging tasks. To address these challenges, we employed a CGAN based on the Wasserstein GAN (WGAN) framework~\citep{wgan}.

Unlike the discriminator in a conventional GAN, which outputs a probability value indicating the confidence in the realness of the sample, the discriminator in a WGAN assigns a score based on the Wasserstein distance (also known as the Earth Mover's distance) between the real and fake distributions. This distance measures the divergence between two probability distributions. Consequently, the discriminator in a WGAN is referred to as a critic. 

Our WGAN model, named $\nu$GAN, consists of a generator and a critic. The generator includes one linear layer and six transposed convolutional layers. The neutrino mass $\sum m_\nu$ is concatenated to the latent vector $z$
of size 200 at the beginning and processed by a linear layer. This is followed by upsampling through six transposed convolutional layers with filter sizes of 5 and 3, and strides of 2 and 1, respectively. Each upsampling operation is followed by batch normalization and a ReLU activation, except for the final layer, which uses a tanh activation without batch normalization. The architecture of the generator is detailed in Table~\ref{tab:wgan_G}.

\begin{table}[hbt]
  \caption{Generator architecture of our $\nu$GAN. \textit{bs} refers to the batch size.}
  \begin{center}
    \begin{tabular}{llll}
      
      \hline
      \multicolumn{1}{c}{\textbf{Layer}} & 
      \multicolumn{1}{c}{\textbf{Operations}} &
      \multicolumn{1}{c}{\textbf{Filter}} &
      \multicolumn{1}{c}{\textbf{Dimension}}\\
      \hline
      
      $z$ &  &  & $\text{bs} \times 200$ \\
      
      $h0$ & linear + identity & & $\text{bs} \times 512 \times 16 \times 16$ \\
      
      $h1$ & deconv + BatchNorm + ReLU & $5 \times 5$ & $\text{bs} \times 256 \times 32 \times 32$ \\
      
      $h2$ & deconv + BatchNorm + ReLU & $5 \times 5$ & $\text{bs} \times 128 \times 64 \times 64$ \\
       
      $h3$ & deconv + BatchNorm + ReLU & $3 \times 3$ & $\text{bs} \times 128 \times 64 \times 64$ \\
      
      $h4$ & deconv + BatchNorm + ReLU & $5 \times 5$ & $\text{bs} \times 64 \times 128 \times 128$ \\
      
      $h5$ & deconv + BatchNorm + ReLU & $3 \times 3$ & $\text{bs} \times 64 \times 128 \times 128$ \\
      
      $h6$ & deconv + Tanh & $5 \times 5$ & $\text{bs} \times 1 \times 256 \times 256$ \\
      
      \hline
    \end{tabular}
    \label{tab:wgan_G}
  \end{center}
\end{table}

The discriminator in our model is composed of four convolutional layers followed by one linear layer. Each convolutional layer uses a filter of size 5 and a stride of 2, doubling the number of channels and halving the feature size with each operation. These layers are followed by batch normalization and a leaky ReLU activation function with a parameter of 0.2. After the convolutional operations, the data is flattened, and the neutrino mass is concatenated to it. Finally, a linear layer is applied to produce the desired output. The detailed architecture of the discriminator is presented in Table \ref{tab:wgan_D}. 

\begin{table}[hbt]
  \caption{Discriminator architecture of our $\nu$GAN. \textit{bs} refers to the batch size.}
  \begin{center}
    \begin{tabular}{llll}
      
      \hline
      \multicolumn{1}{c}{\textbf{Layer}} & 
      \multicolumn{1}{c}{\textbf{Operations}} &
      \multicolumn{1}{c}{\textbf{Filter}} &
      \multicolumn{1}{c}{\textbf{Dimension}}\\
      \hline
      
      $X$ &  &  & $bs \times 1 \times 256 \times 256$ \\
      
      $h0$ & conv + BatchNorm + Leaky ReLU & $5 \times 5$ & $bs \times 64 \times 128 \times 128$ \\
      
      $h1$ & conv + BatchNorm + Leaky ReLU & $5 \times 5$ & $bs \times 128 \times 64 \times 64$ \\
       
      $h2$ & conv + BatchNorm + Leaky ReLU & $5 \times 5$ & $bs \times 256 \times 32 \times 32$ \\
      
      $h3$ & conv + BatchNorm + Leaky ReLU & $5 \times 5$ & $bs \times 512 \times 16 \times 16$ \\
      
      $h4$ & linear + identity & & $bs \times 1$ \\      
      
      \hline
    \end{tabular}
    \label{tab:wgan_D}
  \end{center}
\end{table}

The networks were trained until convergence, defined by a stable distance between the generated and real images, was achieved. The hyperparameters used for training are detailed in Table~\ref{tab:hyperparams_cgan}.

\begin{table}[hbt]
  \caption{Hyperparameters used for model training.}
  \begin{center}
    \begin{tabular}{ll}
      
      \hline
      \multicolumn{1}{c}{\textbf{Hyperparameter}} & 
      \multicolumn{1}{c}{\textbf{Value}}\\
      \hline
      Learning rate (G, D) & $(10^{-5}, 10^{-5})$ \\
      Batch size & $16$ \\
      Latent vector size & $200$ \\
      Latent vector distribution & Standard Normal \\
      Optimizer & ADAM \\
      Gradient Penalty & $1000$ \\
      $\beta_1, \beta_2$ & $0.5, 0.999$\\
      Augmentation & True \\
      Epochs & 300 \\
       
      \hline
    \end{tabular}
    \label{tab:hyperparams_cgan}
  \end{center}
\end{table}

\section{Results} \label{sec:results_cgan}

In this section, we evaluate the performance of our model through both qualitative and quantitative diagnostics. We quantitatively assess the results using a range of summary statistics: power spectrum and transfer function, pixel intensity histogram, pixel peak histogram, and the Multi-Scale Structural Similarity Index to evaluate mode collapse. We also present a visual comparison between the synthetic and real images.

\subsection{Power Spectrum and Transfer Function} \label{subsec:cgan_pow_spec}

\begin{figure}[t]
  \begin{center}
    \includegraphics[width=1.00\textwidth, angle=0]{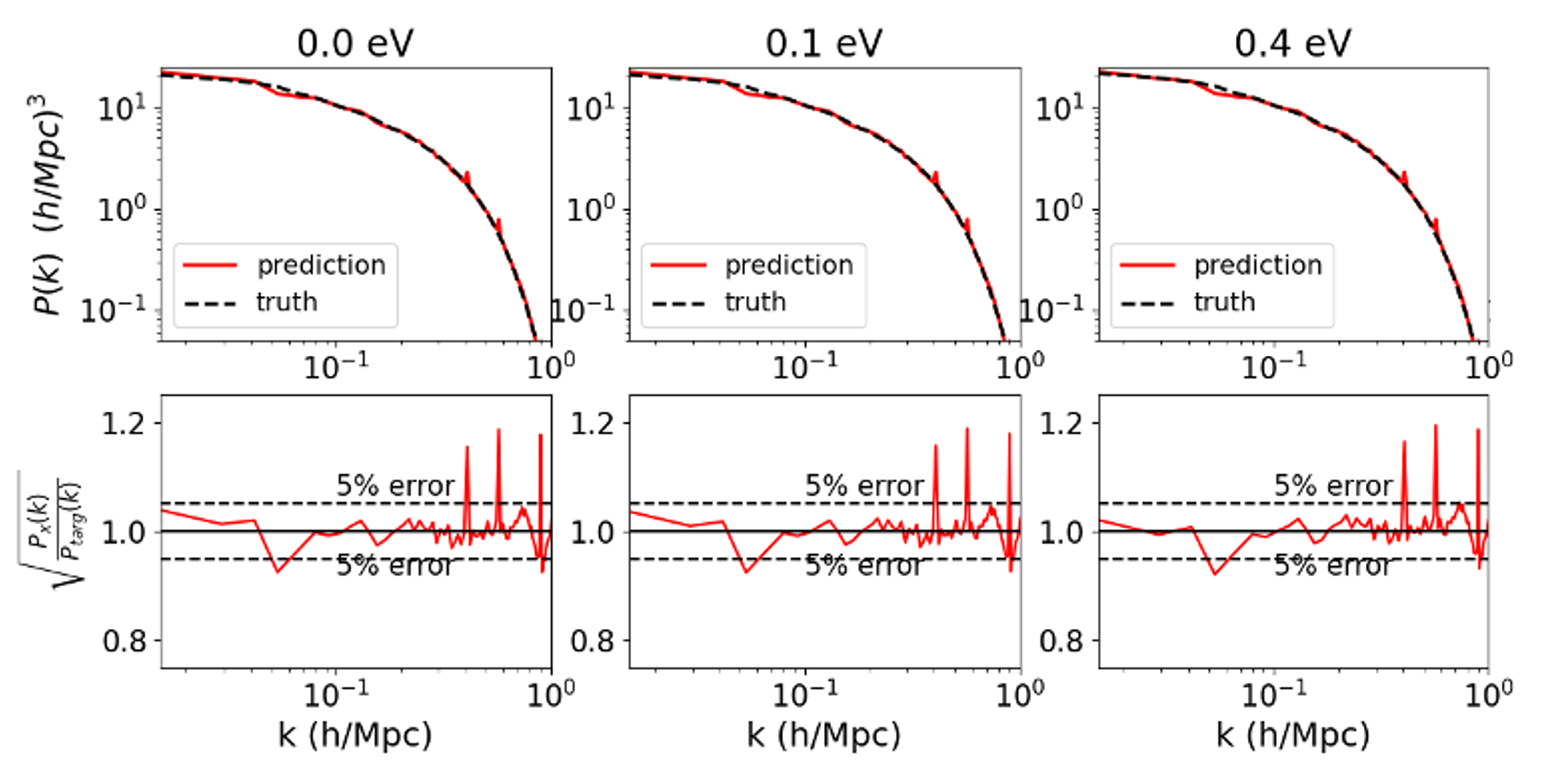}
  \end{center}
  \caption{\textbf{Power spectrum and transfer function comparison.} The top panel shows the average 2D power spectra of the N-body images (black curves) and the ones generated by $\nu$GAN (red curves) for various neutrino masses. The difference in power spectra is very small (within 5\%) at linear and mildly non-linear scales.}
  \label{fig:cgan_pk}
\end{figure}

Figure~\ref{fig:cgan_pk} (top panels) presents the average 2D power spectra computed over 500 realizations, comparing MG-PICOLA simulations (black curve) with the $\nu$GAN model (red curve) for different neutrino masses. The power spectrum analysis provides a quantitative assessment of how well $\nu$GAN captures the statistical properties of large-scale structure formation across various scales. The results demonstrate that the density distribution from $\nu$GAN closely matches the N-body simulations in both amplitude and shape across nearly all scales, particularly at large scales where the agreement is strongest.

To better quantify the agreement between the two datasets, the bottom panels of Figure~\ref{fig:cgan_pk} display the transfer function, which compares the relative clustering amplitude of $\nu$GAN-generated maps and N-body simulation as a function of the wavenumber.~\footnote{The transfer function $T(k)$ is defined as the square root of the ratio of the power spectra of the density field predicted by the $\nu$GAN and the target respectively, $T(k)=\sqrt{{P_{\rm pred}(k)}/{P_{\rm target}(k)}}$} The transfer function should ideally be equal to one if the emulator perfectly reproduces the simulation data. Between $k = 0.01$ and $k = 0.65$, $\nu$GAN achieves an accuracy within 5\%. However, for larger $k$ ($k > 0.7$~\hmpc), deviations become more pronounced. The transfer function shows increasing fluctuations and variance at these smaller scales, indicating the model’s limited ability to capture non-linear clustering due to the lower resolution of the MG-PICOLA training data.

Nevertheless, these preliminary results demonstrate that the $\nu$GAN model outperforms the previous work of ~\cite{elena_giusarma} on mildly non-linear scales ($0.3<k<0.65$\hmpc), achieving an improved accuracy of 5\% compared to the 7\% reported in our previous study, despite being trained on lower-resolution MG-PICOLA simulations with 2D data. Given that our earlier work used high-resolution 3D simulations~\citep{Quijote}, this suggests that incorporating high-resolution 3D data in future work could further enhance the model’s accuracy and reliability.

We note that the power spectrum and transfer function results are averaged over 500 realizations, providing a robust statistical estimate and reducing sample variance.

A key advantage of this approach is that, once trained, the model can generate new cosmic web realizations for different neutrino masses within seconds on a modern Graphics Processing Unit (GPU). This represents a substantial improvement over traditional N-body simulations, which require several hundred to thousands of core-hours per neutrino mass case. As a result, our method provides a gain of several orders of magnitude in computational efficiency, enabling rapid and scalable cosmological emulation.

\subsection{Visual Comparison}\label{subsec:cgan_visual}

\begin{figure}[t]
  \begin{center}
    \includegraphics[width=0.95\textwidth, angle=0]{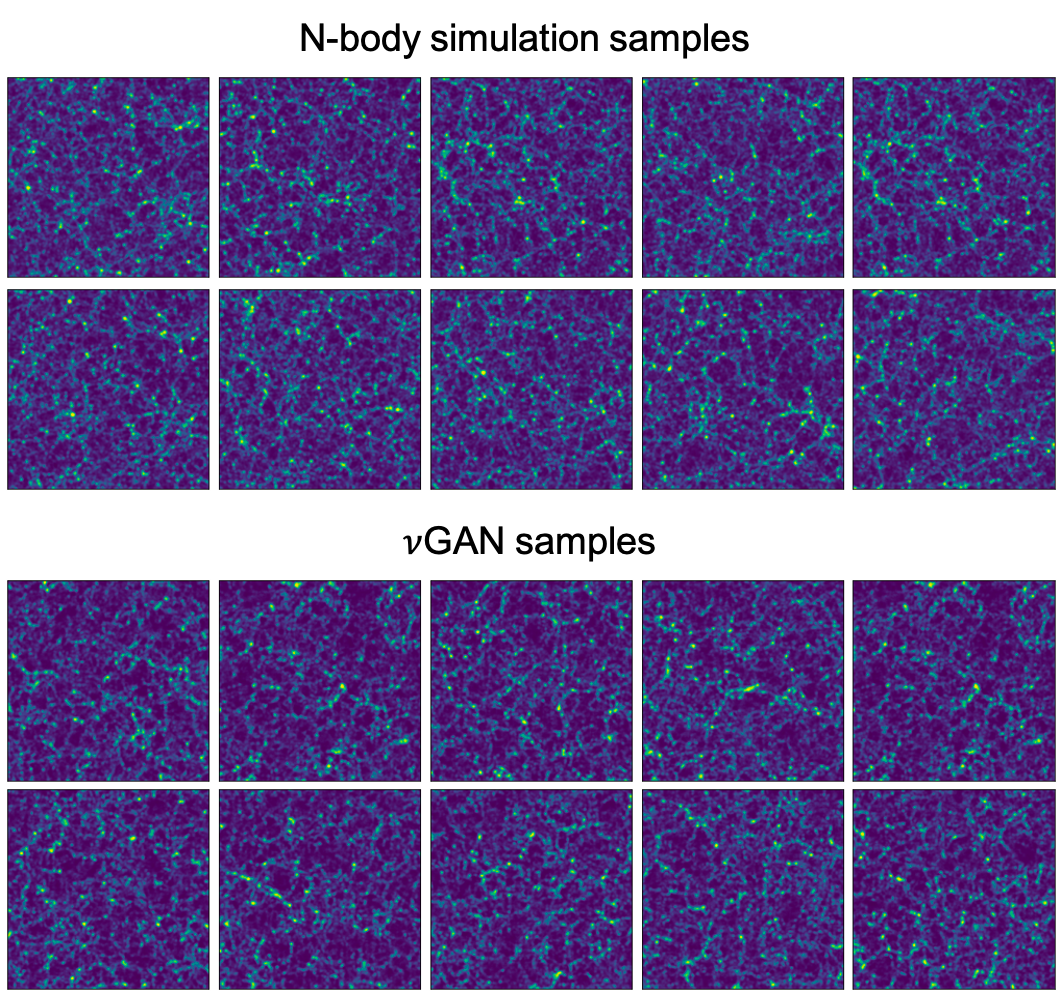}
  \end{center}
  \caption{The top two panels display 10 cosmic web images from N-body simulations, while the bottom two panels show images generated by $\nu$GAN. Each bright spot in the images represents the average number of dark matter particles or the density contrast at that pixel location. It is important to note that the pixel values are scaled to the range of [-1,1].}
  \label{fig:cgan_random_vec_plots}
\end{figure}

Figure \ref{fig:cgan_random_vec_plots} presents $10$ original images of the cosmic web from standard N-body simulations (top) and $10$ synthesized samples generated by $\nu$GAN without neutrinos (bottom) over a scale of $500$ Mpc/h. It is evident that $\nu$GAN effectively captures the prominent visual features of the data. The structures of filaments and halos are well reproduced, demonstrating the capability of GANs to accurately replicate the cosmic web. In particular, the cosmic web structures produced by the our model are visually indistinguishable from the originals.  

\begin{figure}[t]
  \begin{center}
    \includegraphics[width=0.95\textwidth, angle=0]{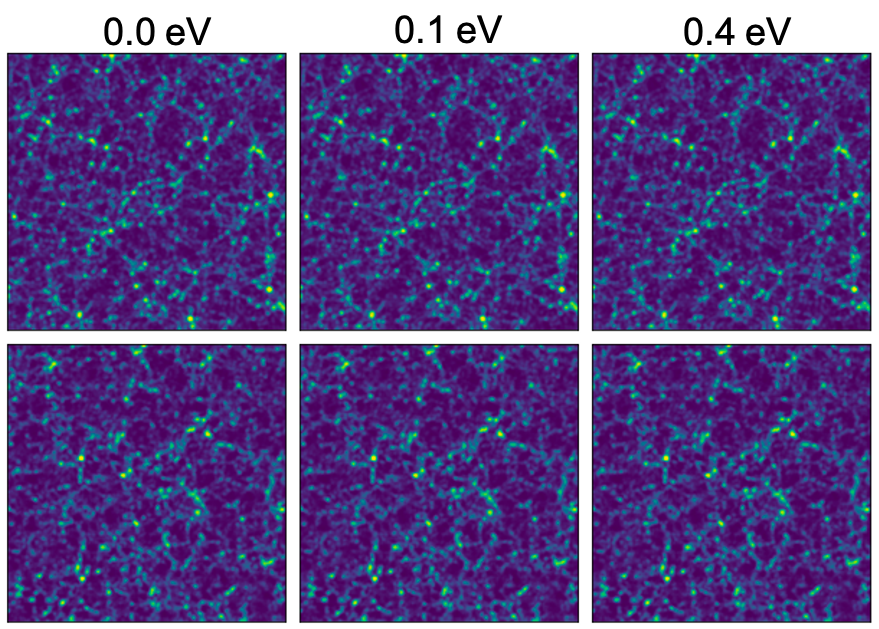}
  \end{center}
  \caption{The top panel displays the cosmic webs from our simulations, while the bottom panel shows those generated by 
$\nu$GAN. In the top images, the random seed was fixed during the simulations, and in the bottom images, the latent vector was fixed. As a result, the images in the top row appear similar to each other, and the same consistency is observed in the bottom row.}
  \label{fig:cgan_single_vec_plots}
\end{figure}

To validate the parameterization of our model on neutrino masses, we generate images using the same latent vector space but with different neutrino masses. Figure  \ref{fig:cgan_single_vec_plots} displays images from N-body simulations (top) and $\nu$GAN (bottom) using various neutrino masses. The latent vector in $\nu$GAN and the random seed in the simulations are kept constant to facilitate a direct visual comparison of the images under the same fixed conditions. Figure~\ref{fig:cgan_single_vec_plots} appears visually very similar across different neutrino masses, reflecting that the impact of neutrino mass on small-scale clustering is too subtle to be visually discernible in 2D cosmic web images, even though the model encodes this dependence in its statistical structure.

\subsection{Pixel Intensity Histogram} \label{subsec:cgan_pixel_intensity}

\begin{figure}[t]
  \begin{center}
    \includegraphics[width=0.95\textwidth, angle=0]{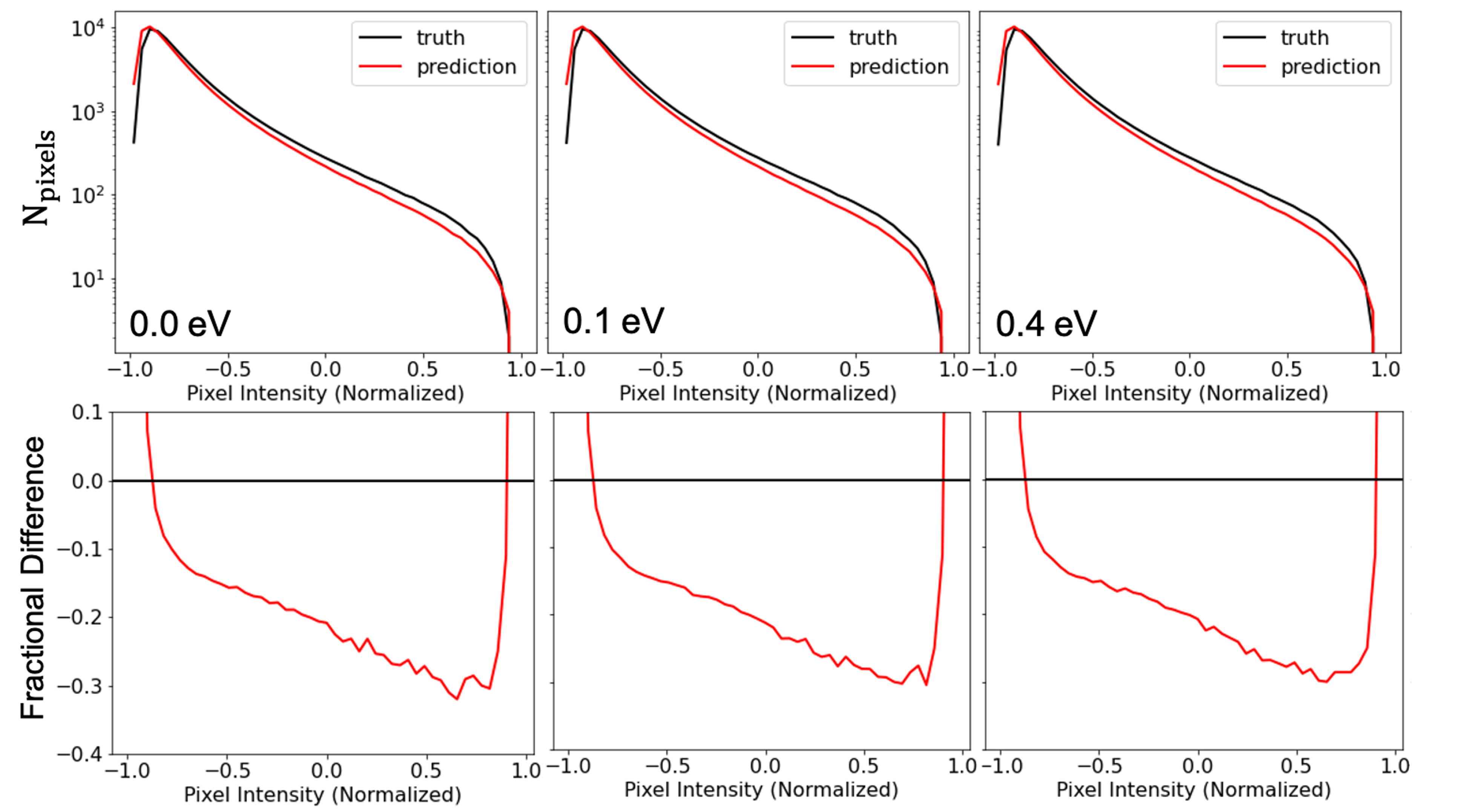}
  \end{center}
  \caption{A comparison of pixel intensity histogram of the samples generated from the N-body simulations and our $\nu$GAN model. The curves are averaged over $500$ samples. The major difference is at lower pixel intensity values.}
  \label{fig:pixel_hist}
\end{figure}

We analyze pixel intensity histograms as complementary diagnostics for assessing $\nu$GAN's ability to replicate cosmic structure. Specifically, we generate mass map histograms, which represent the distribution of mass densities in cosmic web images. Since pixel values in these images correspond to mass density, mass map histograms are also referred to as pixel intensity histograms~\citep{desy3.1, desy3.2}. By comparing these histograms for $\nu$GAN-generated images and N-body simulations, we assess how accurately our model reproduces the true mass distribution of the Universe. Figure~\ref{fig:pixel_hist} illustrates the distribution of mass map pixels ($N_{pixels}$) for both N-body and $\nu$GAN-generated maps. The top panel shows the mass histograms, while the bottom panel presents the fractional difference between the predictions and the actual values. The comparison reveals a generally good agreement between the truth (black curve) and the $\nu$GAN prediction (red curve).
However, for higher pixel intensity values (between $0.25$ and $0.8$), $\nu$GAN predicts a slightly lower number of pixels than the actual N-body samples. The bottom panel indicates that the largest deviations occur at these higher intensity values, with discrepancies of approximately $20\%-30\%$. While these deviations are moderate, their potential impact on cosmological parameter inference should be assessed in future work to determine whether additional corrections or improved training are necessary for precision analyses. Similar trends have been observed in previous works of~\cite{gan_winther} and~\cite{gan_mustafa}.

\subsection{Pixel Peak Histogram} \label{subsec:cgan_pixel_peak_hist}

While the power spectrum provides a complete statistical description of a Gaussian random field, the cosmic web exhibits significant non-Gaussian features that require higher-order statistics for a more detailed characterization. Traditional higher-order statistics, such as the bispectrum and three-point correlation function, provide valuable insights into non-linear structure formation but are computationally expensive to evaluate~\citep{slepian1, slepian2, slepian3, philcox1, philcox2, samushia}. To address these challenges, an alternative approach known as ``peak statistics'' is widely used to extract non-Gaussian features from cosmic web data. Peak statistics have been particularly effective in analyzing weak lensing convergence maps and large-scale structure distributions~\citep{martinet2017,desy3.2}. 

In this context, a peak is defined as a pixel in the mass map that has a greater intensity than all of its surrounding pixels. Specifically, a pixel is classified as a peak if its value exceeds that of all 24 neighboring pixels within a $5\times5$ region. By identifying and quantifying these peaks, we gain additional insight into the clustering of matter, as peaks often correspond to high-density regions such as galaxy clusters.

\begin{figure}[t]
  \begin{center}
    \includegraphics[width=0.95\textwidth, angle=0]{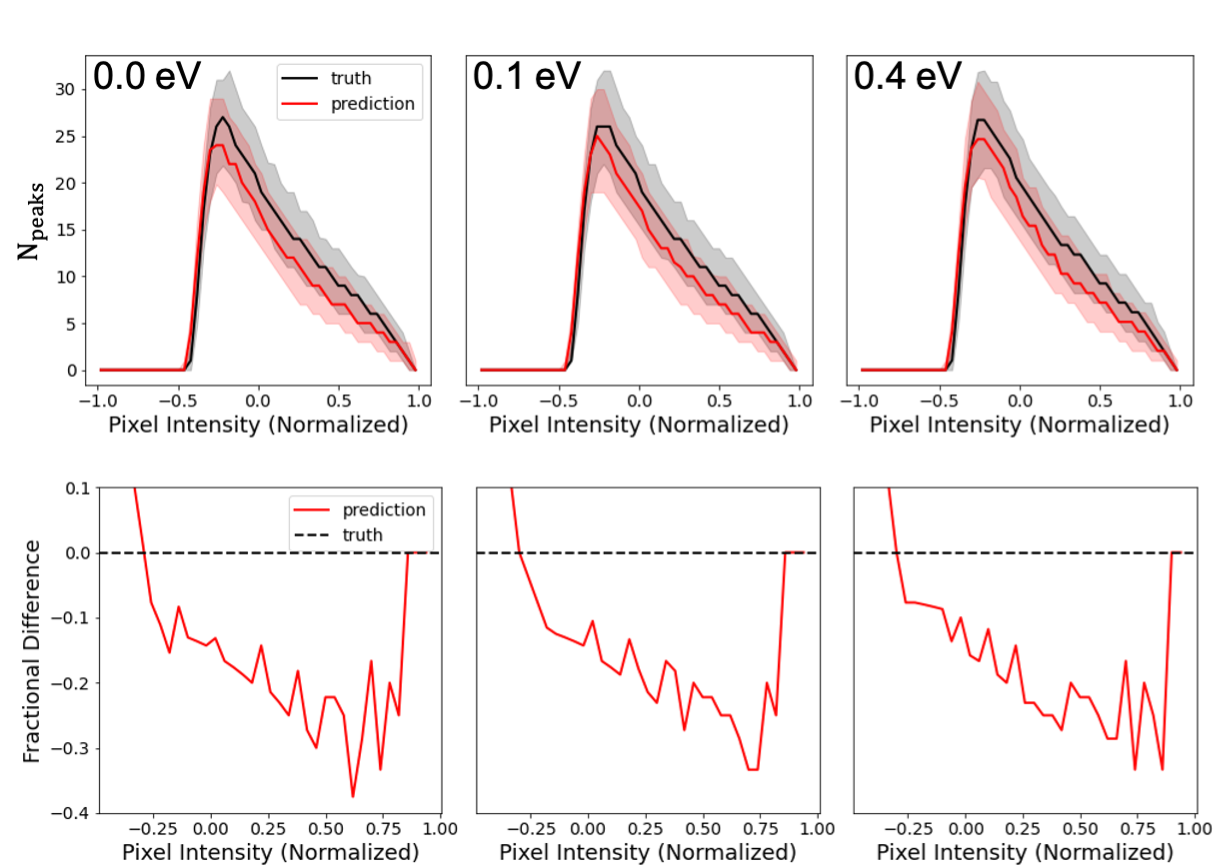}
  \end{center}
  \caption{A comparison of pixel peaks. The solid lines represent the median histogram derived from 500 samples generated by $\nu$GAN and N-body simulations. The shaded areas around each line indicate the $16$th and $84$th percentiles of the distribution. Note that pixel values have been scaled to the range [-1, 1].}
  \label{fig:peak_pixel_histogram}
\end{figure}

Figure~\ref{fig:peak_pixel_histogram} presents the distribution of peaks (denoted as $N_{peaks}$) in the mass map, capturing the local maxima across the dataset. To ensure statistical robustness, this process is applied to $500$ independent simulated images, from which we compute the median histogram of peak counts along with the $16$th and $84$th percentile intervals, capturing the spread of peak values.

The top panel of Figure~\ref{fig:peak_pixel_histogram} displays the histograms of peak counts, while the bottom panel illustrates the fractional difference in peak distributions between the $\nu$GAN-generated images and the N-body simulation results. The overall close agreement between the two datasets demonstrates that $\nu$GAN accurately reproduces the number and distribution of peaks. However, small deviations at the highest peak values suggest that the model may slightly underestimate the frequency of the most massive structures, potentially indicating limitations in capturing the extreme non-linear regime of structure formation.

\subsection{Multi-Scale Structural Similarity Index (MS-SSIM)} \label{subsec:cgan_mssim}

A common challenge with GANs is that, if poorly trained, the generator may repeatedly produce a limited subset of the data distribution, failing to capture the full data variance. This phenomenon, known as \textbf{mode collapse}, results in outputs that represent only a narrow range of the target data\footnote{For example, a GAN trained on a dataset of digits from 0 to 9 might only produce a few digits (e.g., 2, 4, and 9) if it suffers from mode collapse.}.
Our $\nu$GAN avoids mode collapse by using a larger latent vector ($z$) of size 200 from a standard gaussian distribution. Evidence of this can be seen in Figure \ref{fig:cgan_random_vec_plots}, where each generated image is distinctly different, indicating that the model captures a wide range of data features.

To further assess mode collapse, we employ the Multi-scale Structural Similarity Index (MS-SSIM), a widely used metric in image analysis. MS-SSIM helps detect mode collapse, especially relevant in cosmology, where summary statistics may still match true data distributions even when mode collapse is present, potentially concealing the issue.

The MS-SSIM score between two images ranges from $0$ to $1$, where $1$ indicates identical images and $0$ represents completely different images. As the mass maps are stochastic and only statistically similar, we are not interested in similarity between specific maps. Following ~\cite{gan_perraudin1}, we calculate the MS-SSIM between a large ensemble of $2,000$ images from N-body simulations and generated by $\nu$GAN. This approach allows us to evaluate overall similarity across an image set rather than focusing on pairwise comparisons.
Table~\ref{tab:ssim} reports MS-SSIM scores between N-body and $\nu$GAN-generated maps for different neutrino masses. The scores range from 0.41 to 0.58, indicating moderate structural similarity between the generated and real samples. This level of similarity reflects that $\nu$GAN does not replicate individual maps pixel-by-pixel (as expected for stochastic realizations), but rather captures the overall statistical structure of the cosmic web. Importantly, the MS-SSIM values remain sufficiently low to avoid mode collapse, confirming that the model generates diverse samples across the data distribution.

\begin{table}[t]
  \caption{MS-SSIM scores for $\nu$GAN for each neutrino mass}
  \begin{center}
    \begin{tabular}{ll}
      
      \hline
      \multicolumn{1}{c}{\textbf{Neutrino mass (in eV)}} & 
      \multicolumn{1}{c}{\textbf{MS-SSIM score}}\\
      \hline
      $0.0$ & $0.58$ \\
      $0.1$ & $0.48$ \\
      $0.4$ & $0.41$ \\    
      \hline
    \end{tabular}
    \label{tab:ssim}
  \end{center}
\end{table}

\section{Conclusions} \label{sec:conclusions_and_discussions_cgan}

In this work, we demonstrated how GANs can be employed to model the dark matter cosmic web of the Universe, incorporating the effects of neutrinos. We developed a deep Wasserstein GAN, termed $\nu$GAN, which conditions on neutrino masses to generate statistically independent and uncorrelated samples of the 2D cosmic web. By using a larger latent dimension, $\nu$GAN effectively avoids mode collapse, as confirmed by the multi-scale structural similarity (MS-SSIM) scores between generated and real samples. Our model converges efficiently, producing samples that are visually indistinguishable from N-body simulation data.

To validate $\nu$GAN’s accuracy, we evaluated various cosmological and computer vision statistics, including the power spectrum, transfer function, pixel intensity histogram, peak statistics, and MS-SSIM scores, all of which show strong agreement between $\nu$GAN-generated samples and N-body simulations up to mildly non-linear scales. This indicates that 
$\nu$GAN produces genuinely novel data, rather than simply replicating the training set, as observed in previous studies \citep{gan_winther, gan_mustafa, Rodr_guez_2018, gan_perraudin1, gan, gan_feder}.

A key advantage of this approach is that, once trained, the model can generate new cosmic web realizations for different neutrino masses within seconds on a modern Graphics Processing Unit (GPU). This represents a substantial improvement over traditional N-body simulations, which require several hundred to thousands of core-hours per neutrino mass case. As a result, our method provides a gain of several orders of magnitude in computational efficiency, enabling rapid and scalable cosmological emulation. Given the data demands of upcoming large-scale cosmological surveys, efficient methods like $\nu$GAN will be crucial for generating theoretical predictions, facilitating advanced analyses using deep learning \citep{jorit} and sophisticated statistical methods \citep{petri}.

\textit{This work represents a proof-of-concept toward developing rapid, differentiable emulators of the neutrino cosmic web, demonstrating the feasibility and potential of GAN-based approaches for accelerating cosmological simulation pipelines.}
Beyond technical validation, the cosmological implications of our results warrant discussion. The $\sim$5\% accuracy achieved in the power spectrum on linear and mildly nonlinear scales suggests that $\nu$GAN can already serve as a fast emulator for summary statistics relevant to upcoming surveys such as DESI, Euclid, and LSST, especially for large-scale analyses targeting BAO and growth rates. However, the observed deviations of $\sim$20–30\% in peak statistics at high intensities indicate that caution is required when using $\nu$GAN outputs to infer cosmological parameters sensitive to small-scale nonlinearities, such as $\sigma_8$ or $\sum m_\nu$. Future work should quantify how these discrepancies propagate into parameter uncertainties via forward modeling pipelines or likelihood analyses. Nevertheless, this proof-of-concept highlights the potential of generative models to accelerate cosmological inference while retaining acceptable accuracy for many applications.

\section{Future Works} \label{sec:future_works}

While our results demonstrate that $\nu$GAN effectively captures the large-scale statistical properties of the cosmic web, several avenues remain for improvement. One key limitation of the current model is its restriction to 2D projections of the cosmic web. Expanding the framework to generate full 3D density fields is a natural next step, enabling more direct comparisons with state-of-the-art numerical simulations and observational datasets. This will require adapting 3D generative architectures and optimizing their scalability for high-dimensional data.

Additionally, while GANs provide fast sample generation, they suffer from inherent drawbacks such as mode collapse and difficulties in learning the true probability distribution of the data. To address these issues, our ongoing research explores the integration of normalizing flows~\citep{NF} and diffusion models~\citep{ddpm}. Normalizing flows explicitly learn the data probability distribution and provide exact likelihood estimates, making them well-suited for cosmological emulation~\citep{Sultan_2019, sultan2, Lovell:2023gk}. Diffusion models, on the other hand, have demonstrated superior performance in high-resolution image generation and improved mode coverage compared to $\nu$GAN~\citep{mudur2, Mudur, Sether:2024ltp}. Implementing diffusion-based emulators for cosmology could further enhance the accuracy and reliability of deep generative methods in structure formation modeling.

\textit{Future work will focus on improving accuracy at non-linear scales by incorporating high-resolution 3D simulations and leveraging diffusion-based models, thereby extending the applicability of generative emulators to precision cosmology.} 
 While our approach shows promising results, it represents an initial step toward developing more sophisticated generative models, such as diffusion models and normalizing flows, which have the potential to further improve accuracy at non-linear scales.

Furthermore, given that the current approach is trained on low-resolution MG-PCOLA simulations, we plan to incorporate high-resolution N-body simulations such as QUIJOTE~\citep{Quijote} to improve small-scale accuracy and reduce discrepancies in peak statistics. Extending $\nu$GAN to generate simulations across a broader range of cosmological parameters, including baryonic effects, will further increase its applicability to next-generation surveys.

\section{Acknowledgments} \label{sec:ack}
We thank Francisco Villaescusa--Navarro for valuable discussions. NK, EG, and MR acknowledge the IT department at Michigan Technological University for their assistance in managing the computing cluster.

\bibliography{References}{}
\bibliographystyle{aasjournal}

\end{document}